\documentclass[twocolumn,english,aps,prl,reprint,nofootinbib,superscriptaddress]{revtex4-1}
\usepackage[T1]{fontenc}
\usepackage[latin9]{inputenc}
\usepackage{amsmath}
\usepackage{amssymb}
\usepackage{graphicx}

\usepackage{color}
\usepackage{float}
\usepackage{hyperref}

\hypersetup{
    colorlinks,
    linkcolor={black},
    citecolor={black},
    urlcolor={blue}
}

\makeatletter
\@ifundefined{textcolor}{}
{%
 \definecolor{BLACK}{gray}{0}
 \definecolor{WHITE}{gray}{1}
 \definecolor{RED}{rgb}{1,0,0}
 \definecolor{GREEN}{rgb}{0,1,0}
 \definecolor{BLUE}{rgb}{0,0,1}
 \definecolor{CYAN}{cmyk}{1,0,0,0}
 \definecolor{MAGENTA}{cmyk}{0,1,0,0}
 \definecolor{YELLOW}{cmyk}{0,0,1,0}
}

\newcommand{\red}{\textcolor{black}}

\usepackage{pslatex}

\usepackage{babel}

\makeatother

\begin{document}

\title{1D Magneto-Optical Trap of Polyatomic Molecules}

\author{Louis Baum}

\email{louisbaum@g.harvard.edu}
\affiliation{Harvard-MIT Center for Ultracold Atoms, Cambridge, MA 02138, USA}
\affiliation{Department of Physics, Harvard University, Cambridge, MA 02138, USA}

\author{Nathaniel B. Vilas}
\affiliation{Harvard-MIT Center for Ultracold Atoms, Cambridge, MA 02138, USA}
\affiliation{Department of Physics, Harvard University, Cambridge, MA 02138, USA}

\author{Christian Hallas}
\affiliation{Harvard-MIT Center for Ultracold Atoms, Cambridge, MA 02138, USA}
\affiliation{Department of Physics, Harvard University, Cambridge, MA 02138, USA}

\author{\\Benjamin L. Augenbraun}
\affiliation{Harvard-MIT Center for Ultracold Atoms, Cambridge, MA 02138, USA}
\affiliation{Department of Physics, Harvard University, Cambridge, MA 02138, USA}

\author{Shivam Raval}
\affiliation{Harvard-MIT Center for Ultracold Atoms, Cambridge, MA 02138, USA}
\affiliation{Department of Physics, Harvard University, Cambridge, MA 02138, USA}

\author{Debayan Mitra}
\affiliation{Harvard-MIT Center for Ultracold Atoms, Cambridge, MA 02138, USA}
\affiliation{Department of Physics, Harvard University, Cambridge, MA 02138, USA}

\author{John M. Doyle}
\affiliation{Harvard-MIT Center for Ultracold Atoms, Cambridge, MA 02138, USA}
\affiliation{Department of Physics, Harvard University, Cambridge, MA 02138, USA}

\date{\today}

\begin{abstract}
We demonstrate a 1D magneto-optical trap of the polar free radical calcium monohydroxide (CaOH). A quasi-closed cycling transition is established to scatter $\sim 10^3$ photons per molecule, predominantly limited by interaction time. This enables radiative laser cooling of CaOH while compressing the molecular beam, leading to a significant increase in on-axis beam brightness and reduction in temperature from \red{8.4} mK to \red{1.4} mK.

\end{abstract}
\maketitle

Laser cooling and evaporative cooling are key tools of atomic, molecular, and optical physics that are used to produce ultracold atomic and molecular samples \cite{chu1998nobel,phillips1998nobel}. Ultracold atoms have enabled the study of degenerate quantum gases \cite{ketterle2002nobel}, high-precision clocks \cite{Swallows1043}, quantum many-body physics, and quantum simulation of condensed matter systems \cite{bloch2008many}. Polar molecules, with their additional internal degrees of freedom and long-range interactions, promise further access to novel phenomena in the ultracold regime \cite{carr2009cold}. For example, diatomic molecules have been identified for applications that include precise searches for physics beyond the Standard Model \cite{baron2014order,andreev2018improved,hudson2002measurement,cairncross2017precision}, quantum simulation \cite{micheli2006toolbox,manmana2013topological,Anderegg1156}, studies of fundamental collisional \cite{segev2019collisions} and chemical \cite{bohn2017cold,ospelkaus2010quantum,Hu1111} processes, and production of exotic ultracold atoms through photodissociation of ultracold molecules \cite{C2CP42709E,C1CP21304K}. While diatomic molecules are a rich resource (and are only beginning to be explored), polyatomic molecules have qualitatively distinct advantages at the frontier of quantum science. Ultracold polyatomic molecules have been identified for applications including improved precision searches for the electron EDM \cite{kozyryev2017precision} and for dark matter \cite{kozyryev2018enhanced}, novel quantum computation \cite{tesch2002quantum,wei2011entanglement,yu2019scalable} and quantum simulation platforms \cite{wall2015realizing,wall2013simulating}, the control of the primordial chemical reactions that gave rise to life \cite{lazcano1996origin}, the study of biomolecular chirality \cite{quack1989structure,quack2002important}, and the study of ultracold collisions and quantum chemistry in increasingly complex systems, while maintaining single quantum state control \cite{augustovicova2019collisions}.

With such promise, there have been intense efforts to cool molecules. ``Indirect'' approaches, such as association techniques, like coherent adiabatic binding of laser-cooled atoms, have led to a variety of ultracold diatomic bialkali samples including a quantum degenerate Fermi gas of KRb \cite{de2019degenerate,Ni231}. ``Direct'' cooling approaches use electromagnetic fields (e.g. lasers or pulsed external fields) to slow and cool molecules \cite{fulton2004optical,lavert2011moving,wu2017cryofuge,vanhaecke2007multistage,fitch2016principles,petzold2018zeeman,shuman2009radiative,hemmerling2016laser}. These techniques include Sisyphus-type approaches, which have, for example, produced samples of H$_2$CO as cold as 420 $\mu$K \cite{zeppenfeld2012sisyphus,prehn2016optoelectrical}. Laser cooling has been identified as being potentially applicable to a variety of molecular structures \cite{stuhl2008magneto,di2004laser,carr2009cold,mccarron2018laser,tarbutt2018laser,lim2018laser,truppe2019spectroscopic,C1CP21313J}, including polyatomic species composed of a single metal atom bound to an electronegative radical, called ``MOR'' molecules \cite{kozyryev2016proposal,Ivanov2020,klos2019prospects}. Crucially, laser cooling offers a path to trapped $\mu$K samples of molecules in single internal and motional quantum states. SrF \cite{shuman2009radiative,shuman2010laser,barry2012laser,barry2014magneto,mccarron2015improved,norrgard2016submillikelvin,steinecker2016improved}, CaF \cite{hemmerling2016laser,chae2017one,anderegg2017radio,williams2017characteristics,truppe2017molecules,truppe2017intense,caldwell2019deep}, and YO \cite{hummon20132d,yeo2015rotational,collopy20183d} have all been laser cooled and loaded into magneto-optical traps (MOTs). SrF and CaF have been cooled below the Doppler limit and transfered to optical or magnetic traps \cite{anderegg2018laser,mccarron2018magnetic,williams2018magnetic}. Sisyphus laser cooling of the polyatomic molecules SrOH and YbOH has been achieved \cite{kozyryev2017sisyphus,augenbraun2019laser}, and coherent optical forces have been applied to SrOH \cite{kozyryev2018coherent}. 

\begin{figure}[ht]
\begin{centering}
\includegraphics[width = 0.9\columnwidth]{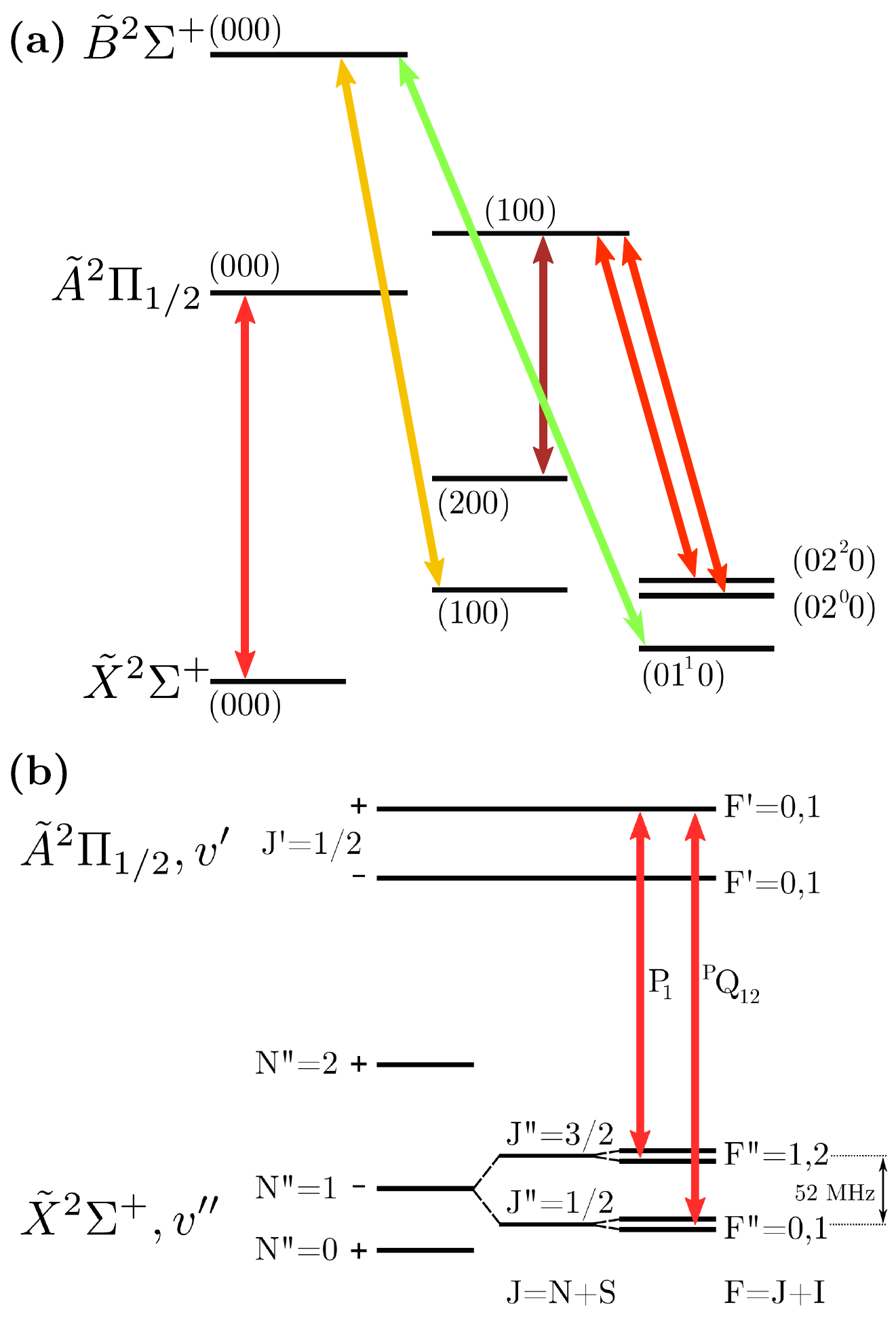} 
\caption{\label{fig:CaOH_Levels} (a) Laser cooling scheme for CaOH. The vibrational structure depicted here indicates all levels that are addressed with lasers in order to limit the branching ratio to other vibrational states to \red{$4.5\times 10^{-4}$}. (b) Rotational structure of CaOH illustrating the 52 MHz spin-rotation splitting in the electronic ground state as well as the unresolved hyperfine structure (1.5 MHz and 7 kHz in the $J''=\frac{3}{2}$ and $J''=\frac{1}{2}$ states respectively \cite{scurlock1993hyperfine}). The $\tilde{X}^2\Sigma^+\left(v_1'' v_2'' v_3''\right) \rightarrow\tilde{A}^2\Pi_{1/2}\left(v_1' v_2' v_3'\right)$ $P_{1}(J''=\frac{3}{2})$ and $^PQ_{12}(J''=\frac{1}{2})$ rotationally closed transitions are shown \cite{di2004laser}. The parity of the ground states is indicated by the sign to the right of the $N''$ value while the parity of the excited states is indicated to the right of the $J'$ value. The rotational structure of the $\tilde{B}^2\Sigma^+(000)$ state is analogous to that of the $\tilde{X}^2\Sigma^+$ states and is not pictured. Rotational closure on repumping lines through this state is achieved by driving $P_1(J''=\frac{3}{2})$ and $^PQ_{12}(J''=\frac{1}{2})$ transitions to the $\tilde{B}^2\Sigma^+(N'=0, J' = \frac{1}{2},+)$ state. The level diagrams are not to scale.}
\par\end{centering}
\end{figure}

\begin{figure*}[ht]
\begin{centering}
\includegraphics[scale = 1]{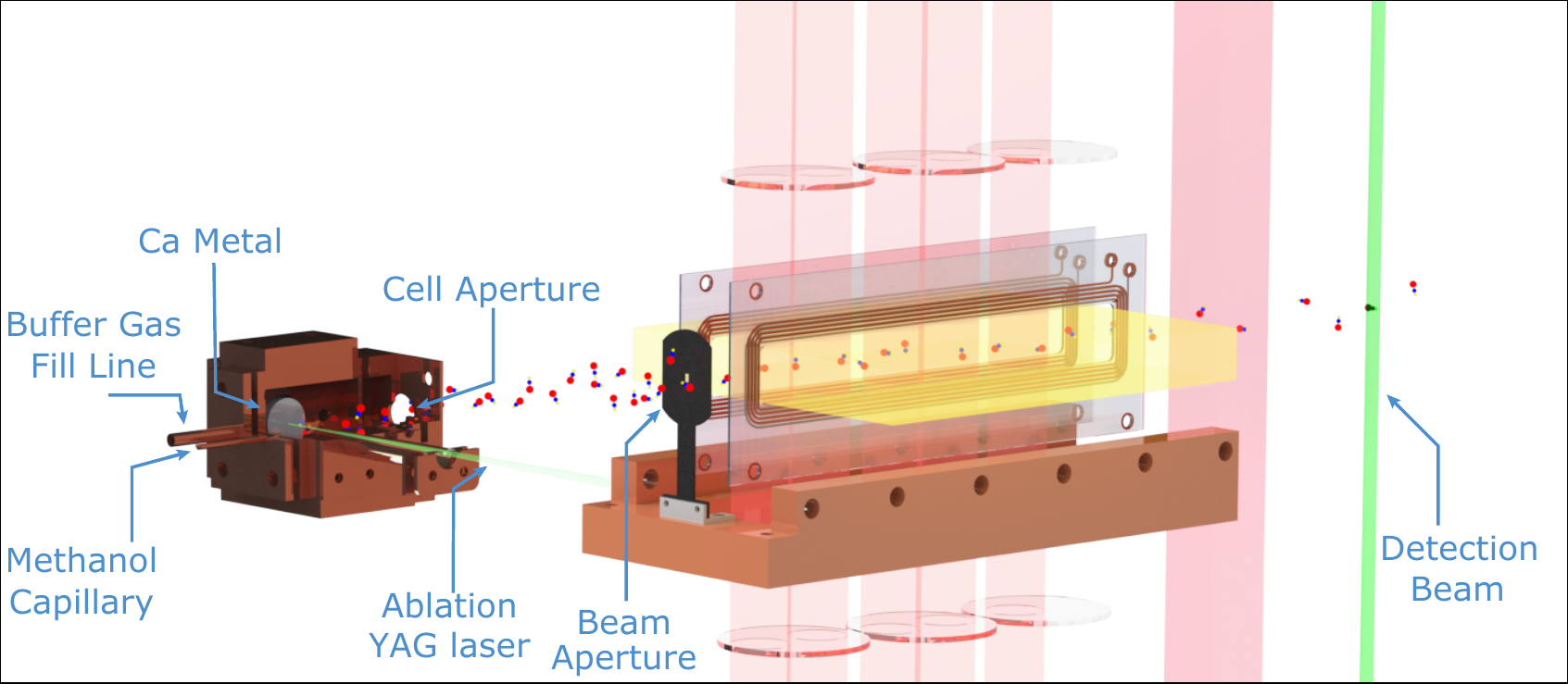}
\caption{\label{fig:Experimental_Diagram} A rendering of the experimental apparatus. On the far left is the two-stage buffer-gas beam source, depicted in cut-away view for clarity. 35.5 cm from the exit of the buffer-gas cell, the molecular beam is collimated by a 3 mm square beam aperture. 39 cm from the cell, the molecules enter the interaction region where they are addressed with light from the main MO cooling beams in the vertical direction. Co-propagating vertically are the $(100), (200)$ and $(02^00)$ repumping lasers. The $(02^20)$ and $(01^10)$ repumping light is multipassed in the horizontal direction and extends beyond the MO region. A separate vertically multipassed region containing $(100)$ and $(02^00)$ repumping light lies after the magnetic field coils and serves to recover population from excited vibrational states. Finally, the molecules encounter a detection beam of smaller cross-section than the cooling and repumping light, and the resulting laser-induced fluorescence is collected and imaged onto an EMCCD.}
\par\end{centering}
\end{figure*}

In this Letter, we demonstrate radio frequency (RF) magneto-optical (MO) cooling and compression (1D MOT) of a beam of the polyatomic molecule $^{40}$Ca$^{16}$OH, an archetypal example of the broader class of MOR molecules. In doing so, we realize a cycling scheme capable of scattering $\sim$ 10$^3$ photons. We characterize the MO forces applied here by extracting force constants and damping rates. A concomitant on-axis increase in molecular density is observed. This demonstration of MO cooling establishes a route towards deep laser cooling and optical trapping for numerous species of polyatomic molecules.

Effective MO cooling and compression requires scattering many photons without losing population to states that do not couple to the laser light (``dark states''). Establishing such a cycling transition in molecules requires closing both vibrational and rotational degrees of freedom, as depicted in Fig \ref{fig:CaOH_Levels}. Vibrational decay is not governed by rigorous selection rules but instead by wavefunction overlap, which is quantified by Franck-Condon factors (FCFs). CaOH is an example of a broad class of polyatomic molecules that have been identified as promising candidates for laser cooling due to their diagonal FCFs and strong electronic transitions \cite{kozyryev2016proposal,kozyryev2019determination}. The main laser cooling transition in CaOH is the $\tilde{X}^2\Sigma^+\left(000\right) \rightarrow\tilde{A}^2\Pi_{1/2}\left(000\right)$ transition with a natural linewidth of 2$\pi$ $\times$ 6.4 MHz at 626 nm \cite{CaOHAlifetime}. The highly diagonal FCFs of the $\tilde{A}^2\Pi_{1/2}\left(000\right)$ state suppress spontaneous decay to higher vibrational states during a single scattering event; nonetheless, significant optical pumping into excited vibrational states can occur when many photons are scattered. CaOH has three vibrational modes: a symmetric stretch, a doubly degenerate bend, and an antisymmetric stretch. These vibrational modes are labeled with four quantum numbers $\left(v_1,{v_2}^l,v_3\right)$, where $v_1$, $v_2$, and $v_3$ indicate the number of quanta in the symmetric stretching mode, the bending mode, and the antisymmetric stretching mode, respectively. $l$ labels the nuclear orbital angular momentum in the bending mode and takes values of $l=-v_2,-v_2+2,...,v_2$ \cite{herzberg1966molecular}. Five repumping lasers, listed in Table \ref{tab:Transitions}, are used to establish a quasi-closed cycling scheme and recover population in these states, as depicted in Fig. \ref{fig:CaOH_Levels}. Branching ratios within this cycling scheme are reported in the Supplemental Material. 

Notably, both the $\tilde{X}^2\Sigma^+\left(01^10\right)$ and $\tilde{X}^2\Sigma^+\left(02^20\right)$ states need to be repumped. Decays to these states are nominally forbidden by an approximate $\Delta l = 0$ selection rule that originates from the separation of electronic and vibrational degrees of freedom in the Born-Oppenheimer approximation. The breakdown of this selection rule has been observed previously for $\Delta l = 1$ transitions in CaOH (and other similar systems) and is attributed to a second order process involving Renner-Teller mixing and spin-orbit coupling leading to intensity borrowing via the $\tilde{B}^2\Sigma^+\left(01^10\right)$ state \cite{brazier1985laser,kozyryev2019determination}. Decay to the $\tilde{X}^2\Sigma^+\left(02^20\right)$ state was previously unobserved. We attribute the magnitude of this decay to a similar mechanism that relies on the mixing of vibrational states within the $\tilde{A}^2\Pi_{1/2}$ manifold (see Supplemental Material). We measure the branching ratio out of this cycling scheme to be $4.5(\red{7}) \times 10^{-4}$, which is predicted to be dominated by decay to the $\tilde{X}^2\Sigma^+\left(12^00\right)$, $\tilde{X}^2\Sigma^+\left(12^20\right)$, and $\tilde{X}^2\Sigma^+\left(300\right)$ vibrational states. Details of this measurement will be the subject of a subsequent publication.

To avoid populating rotational dark states, each laser beam (main and all repumpers) contains two frequency components separated by the spin-rotation (SR) splitting of 52 MHz depicted in Fig \ref{fig:CaOH_Levels} (b). The hyperfine splitting is below the natural linewidth of the main cooling transition and does not require additional frequency sidebands \cite{scurlock1993hyperfine}.  This type of transition ($J \rightarrow J' = J-1$) causes rapid optical pumping into magnetic dark states, significantly reducing the cooling and confining forces in molecular MOTs \cite{tarbutt2015magneto}. We address this by simultaneously switching both the laser polarization and the sign of the magnetic field gradient during cooling, which evolves magnetic dark states into bright states, as previously demonstrated in diatomic systems \cite{anderegg2017radio,norrgard2016submillikelvin,hummon20132d}.

\begin{table}
\centering
\begin{tabular}{| r c l | c |}
\hline
\multicolumn{3}{|c|}{Transition} & Wavelength (nm) \\
\hline
$\tilde{X}^2\Sigma^+\left(000\right)$ & $\rightarrow$ & $\tilde{A}^2\Pi_{1/2}\left(000\right)$ & 626.4 \\
$\tilde{X}^2\Sigma^+\left(100\right)$ & $\rightarrow$ & $\tilde{B}^2\Sigma^+\left(000\right)$ & 574.3 \\
$\tilde{X}^2\Sigma^+\left(200\right)$ & $\rightarrow$ & $\tilde{A}^2\Pi_{1/2}\left(100\right)$ & 650.4 \\
$\tilde{X}^2\Sigma^+\left(02^00\right)$ & $\rightarrow$ & $\tilde{A}^2\Pi_{1/2}\left(100\right)$ & 629.0 \\
$\tilde{X}^2\Sigma^+\left(02^20\right)$ & $\rightarrow$ & $\tilde{A}^2\Pi_{1/2}\left(100\right)$ & 630.0 \\
$\tilde{X}^2\Sigma^+\left(01^10\right)$ & $\rightarrow$ & $\tilde{B}^2\Sigma^+\left(000\right)$ & 566.0 \\
\hline
\end{tabular}
\caption{Optical transitions and corresponding wavelengths driven to form a quasi-closed cycling transition in CaOH. The $\tilde{X}^2\Sigma^+\left(000\right) \rightarrow \tilde{A}^2\Pi_{1/2}\left(000\right)$ transition is the main cooling line while the other five frequencies correspond to vibrational repumping lasers.}
\label{tab:Transitions}
\end{table}

\begin{figure}[t]
\begin{centering}
\includegraphics[width =.45\textwidth]{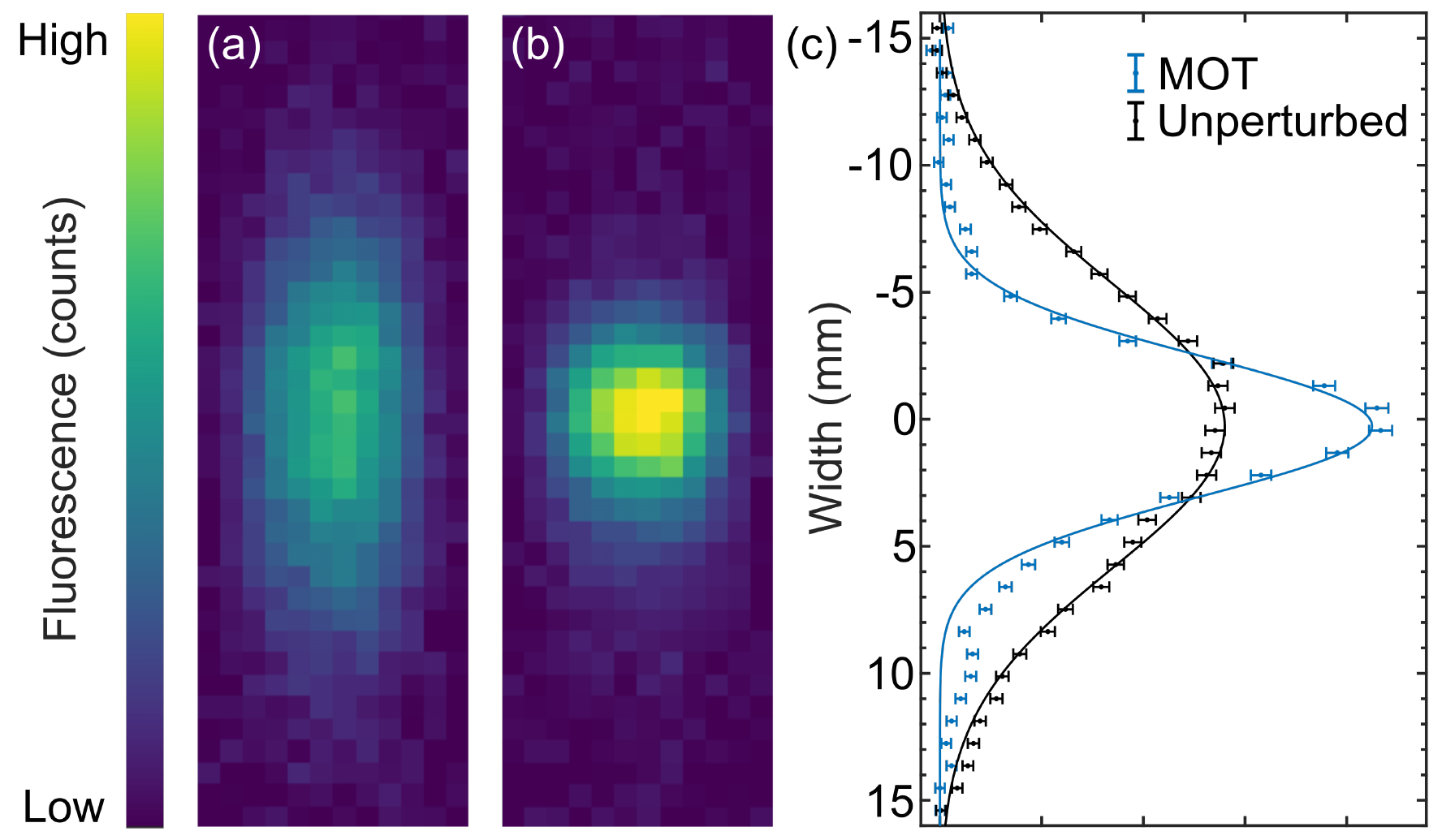}
\par\end{centering}
\caption{\label{beamprofiles} Raw images of the molecular beam taken for (a) unperturbed CaOH molecules and (b) under 1D MOT conditions. The molecular beam propagates from left to right in these images, while the cooling and detection light propagates in the vertical direction. (c) Horizontally integrated molecular beam profiles indicating cooling and on-axis density enhancement. Solid lines are fits to a Gaussian profile. These beam profiles have not been rescaled. The area of the MO compression trace is 78$\%$ of the unperturbed beam. Using the measured branching ratio out of our photon cycling scheme, this corresponds to $\sim$ 550 photons scattered. At this laser intensity, on-axis beam brightness is optimized. At higher intensities, we observe saturation of the cooling and compression effects, while the concomitant additional photon scattering leads to population loss to unaddressed vibrational states.}
\end{figure}

CaOH molecules are produced using a cryogenic buffer gas source \cite{hutzler2012buffer,barry2011bright} as depicted in Fig \ref{fig:Experimental_Diagram}. Hot calcium atoms are produced by laser ablation of a metallic calcium target inside of a copper cell held at $\sim$ 2 K while flowing 6 standard cubic centimeters per minute (SCCM) of helium buffer gas. We simultaneously flow a small amount ($\sim$ 0.01 SCCM) of methanol vapor into the cell through a thermally isolated capillary at $\sim$ 250 K. Methanol molecules react with calcium atoms to produce CaOH. The CaOH molecules rapidly cool via collisions with the helium buffer gas. This produces CaOH at densities of $\sim$ 10$^{10}$ cm$^{-3}$ in a single rotational state, as measured by laser absorption in the cell. The cold CaOH molecules are entrained in the buffer gas flow and extracted from a two-stage cell into a cryogenic buffer-gas beam (CBGB) with a mean forward velocity of $v_f \sim$100 m/s and a transverse velocity spread of $v_{\perp} \sim$20 m/s \cite{hutzler2012buffer}. The CBGB is collimated by a 3 mm square aperture located 35.5 cm from the exit of the buffer-gas cell, resulting in a transverse temperature $T_{\perp}$ $\sim$ \red{8.4 mK}.

After exiting the aperture, the collimated molecular beam enters the interaction region containing six distinct wavelengths of light (main plus five repumpers). The combined laser light, with a beam diameter of 25 mm, makes 5 round trip passes through the interaction region as well as through a pair of $\lambda$/4 waveplates for 12.5 cm of total interaction length. The main laser cooling light is circularly polarized and retroreflected in a $\sigma^+- \sigma^-$ configuration. Details are provided in the Supplemental Material. The handedness of the polarization is rapidly switched using a voltage-variable waveplate (Pockels cell). A quadrupole magnetic field is generated with a pair of in-vacuum anti-Helmholtz coils and sinusoidally driven at the same frequency as the laser polarization switching with a controllable phase offset. 

\begin{figure}[t]
\begin{centering}
\includegraphics[width=.45 \textwidth]{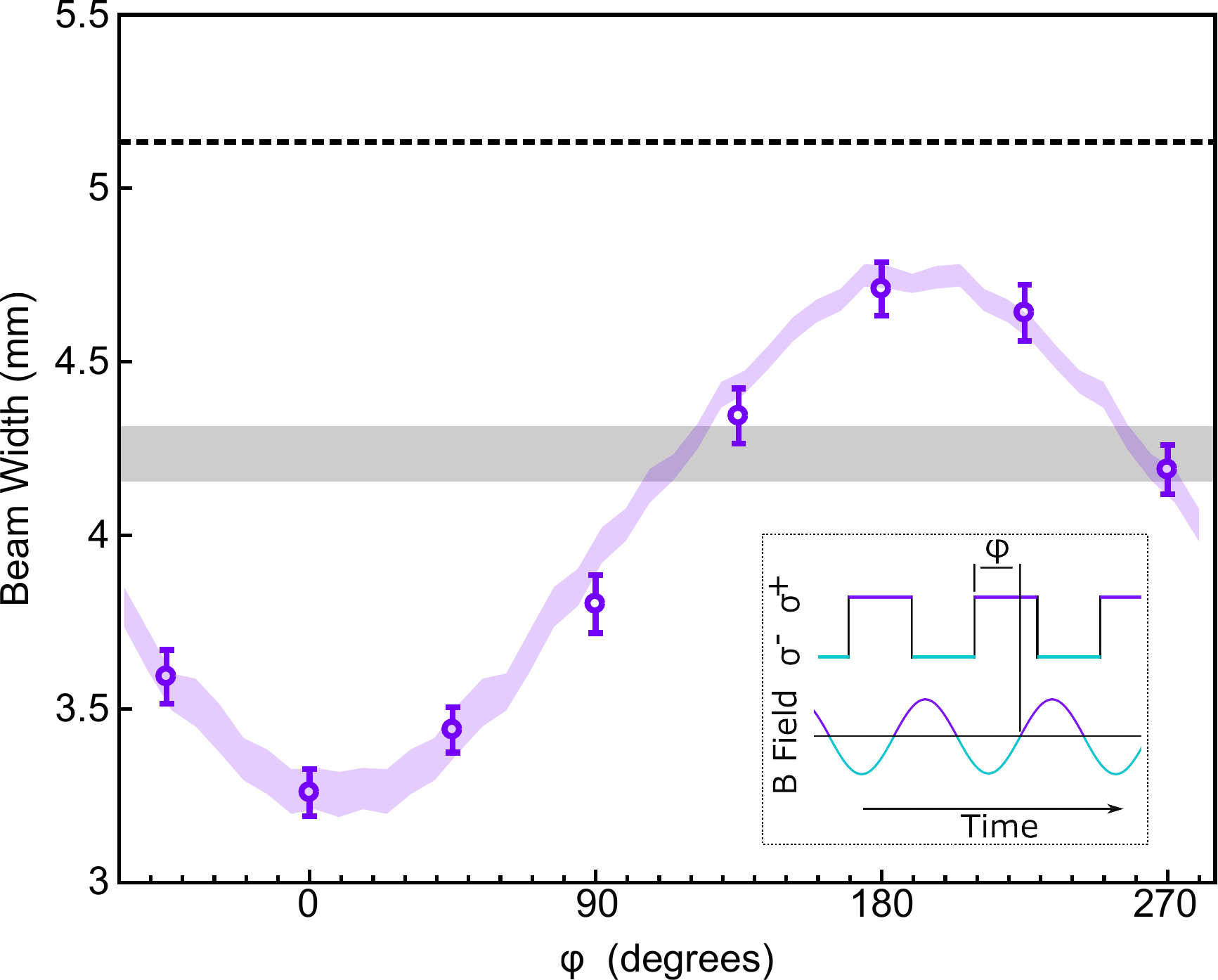}
\par\end{centering}
\protect\caption{\label{phasescan} Molecular beam width as a function of phase offset ($\varphi$) between the polarization switching of cooling light and the oscillating magnetic field (see inset). The dashed black line indicates the width of the unperturbed beam, the shaded grey region indicates the measured width and associated error of the Doppler cooled beam without an applied magnetic field, and the shaded purple region indicates Monte Carlo simulation results for the full MO configuration. Clear compression at 0 phase corresponds to the MOT configuration, where the laser polarization provides a spatially confining force. At 180 degrees there is expansion of the beam, corresponding to the anti-MOT configuration. The main cooling light is detuned -7 MHz from resonance while all repumping lasers remain resonant. Error bars represent one standard deviation of the fitted Gaussian beam width.}
\end{figure}

Following the interaction region, where MO cooling and compression take place, repumping lasers are applied to recover population from excited vibrational states. The molecules expand ballistically while propagating to the detection region, mapping the momentum distribution onto the spatial extent of the molecular beam. The molecules are then excited with lasers addressing the $\tilde{X}^2\Sigma^+\left(000\right) \rightarrow\tilde{B}^2\Sigma^+\left(000\right)$ and $\tilde{X}^2\Sigma^+\left(100\right) \rightarrow\tilde{B}^2\Sigma^+\left(000\right)$ lines with the resulting laser-induced fluorescence imaged onto an EMCCD camera. The collection efficiency of the imaging system is measured to be constant over the region occupied by the molecules. The resulting image is integrated along the direction of molecule propagation to produce a spatial beam profile, which we fit to a Gaussian distribution. We parameterize the width of the molecular beam by the standard deviation of the Gaussian fit. MO cooling and compression are seen as a narrowing of this width, as shown in Fig \ref{beamprofiles}. The main cooling laser intensity was 1.6 mW/cm$^2$ for the data in Fig \ref{beamprofiles} and 3.3 mW/cm$^2$ for the data in Fig \ref{phasescan}. All data were collected with an RF switching frequency of 530 kHz, a detuning of -7 MHz, and an RF voltage applied to the coils corresponding to a root-mean-square magnetic field gradient of 17 Gauss/cm. Further details on the apparatus are contained in the Supplemental Material.

In order to differentiate Doppler and MO effects, we scan the phase of the polarization switching relative to the magnetic field gradient switching, as shown in Fig \ref{phasescan}. The greatest compression of the beam occurs at a phase of 0 degrees and corresponds to the MOT configuration, while at a phase of 180 degrees we see expansion of the beam, corresponding to the anti-MOT. The observed phase dependence is a clear signature of the application of MO forces in addition to the effects of Doppler cooling alone, represented by the gray shaded region in the figure. By measuring the loss of molecules to vibrational dark states as a function of cooling light intensity and by comparing to the known branching ratios of repumped vibrational levels, we are able to determine the number of photons scattered by the cooling process. We find that we can scatter up to \red{920$^{+170}_{-120}$} photons during the cooling process, limited primarily by interaction time. The beam compression saturates after \red{$\sim$ 550 photons} are scattered. We attribute this saturation to a combination of sub-Doppler heating and MO overfocusing of the molecular beam. 

As a means of characterizing our system we use a Monte Carlo simulation to model molecular propagation and cooling dynamics. The MO forces are described by an effective rate-equation model developed previously in diatomic systems \cite{norrgard2016submillikelvin} and described in detail in the Supplemental Material. The resulting forces can be linearized in the form $F_\text{MO}/m \approx -\beta v - \omega^2 r$, where $m$ is the molecular mass, $r$ and $v$ are the position and velocity of the molecules, $\omega$ is the MO oscillation frequency, and $\beta$ is the damping constant. By fitting the results of this model to our data we extract MO cooling parameters $\omega \approx 2\pi \times 90$ Hz and $\beta \approx 400$ s$^{-1}$. These values are comparable to those observed for 2D and 3D MOTs of diatsomic molecules \cite{hummon20132d, mccarron2015improved, anderegg2017radio, truppe2017molecules}. By fitting the final velocity distribution of the molecular cloud after propagation through the simulated cooling region, we extract transverse beam temperatures. After Doppler cooling alone we find \red{$T$ = 3.1(1) mK} (from an initial temperature of \red{$T$ = 8.4(2) mK}); with MO cooling and compression the temperature is further reduced to \red{$T$ = 1.4(1) mK}. The simulated MOT force is then used to extract an on-axis capture velocity of $\sim 7$ m/s for a 3D MOT of CaOH, which is similar to that measured in diatomic molecules \cite{williams2017characteristics}.

In summary, we demonstrate magneto-optical cooling and compression of polyatomic CaOH molecules. We establish a cycling transition and scatter up to $\sim$ 10$^3$ photons, limited primarily by interaction time. We also observe cooling from \red{8.4 mK to 1.4 mK}. This technique could be used as a means of increasing beam brightness to substantially enhance molecule numbers loaded into a 3D MOT. Demonstrating this degree of photon cycling sets the stage for optical slowing of a molecular beam and ultimately the realization of a full 3D MOT. As a result, this work represents a significant step forward in extending cooling and trapping techniques to larger, more complicated molecular species, which will allow the production of ultracold polyatomic molecular samples and deep cooling into the $\mu$K regime.

We would like to thank L. Anderegg for insightful discussions. This work was supported by the NSF. N.B.V. acknowledges support from the NDSEG fellowship, and B.L.A. from the NSF GRFP.

\bibliographystyle{apsrev4-1}
\bibliography{1DMOTreferences}

\pagebreak

\clearpage

\setcounter{equation}{0}

\setcounter{figure}{0}

\renewcommand{\thefigure}{S\arabic{figure}}

\renewcommand{\theequation}{S\arabic{equation}}

\renewcommand{\bibnumfmt}[1]{[#1]}

\renewcommand{\citenumfont}[1]{#1}

\onecolumngrid

\begin{center}

\large{\textbf{Supplemental Material}}
\vspace{.5 cm}

\normalsize
\begin{minipage}{.9\linewidth}
In this Supplement, we discuss the measured vibrational branching ratios out of the laser cooling cycle used in the experiment. We provide further details on the laser polarizations and Zeeman structure used to apply magneto-optical forces to CaOH molecules, as well as on the experimental apparatus used to perform the laser cooling and compression. Finally, we detail the Monte Carlo model used to simulate molecular propagation and cooling dynamics and extract the MO cooling parameters quoted in the main text.

\end{minipage}
\end{center}
\twocolumngrid

\begin{center}
\textbf{Branching Ratios and Scattered Photon Numbers}
\end{center}

The vibrational branching ratios (VBRs) describing vibronic decay from the excited $\tilde{A}^2\Pi_{1/2}$ and $\tilde{B}^2\Sigma^+$ states to various vibrational levels of the ground $\tilde{X}^2\Sigma^+$ electronic state are determined using previously-reported dispersed laser-induced fluorescence (DLIF) \cite{kozyryev2019determination} in combination with molecular beam deflection and optical pumping measurements. They are given in Table \ref{tab:FCFs}. The details of these measurements are beyond the scope of this work and will be the subject of a subsequent publication currently in preparation. Note that Franck-Condon factors and VBRs differ slightly due to the factor of $\nu^3$, where $\nu$ is the transition frequency, when relating wavefunction overlap to transition strength \cite{kozyryev2019determination}. Since our laser cooling scheme involves multiple excited electronic states we are sensitive to decay out of the cooling scheme as a whole rather than from specific excited states.

While decay from the $\tilde{A}^2\Pi_{1/2}(000)$ state dominates most loss channels, decay to the $\tilde{X}^2\Sigma^+(01^10)$ state occurs in part via the $\tilde{B}^2\Sigma^+(000)$ state in our repumping scheme. In particular, because the $\tilde{X}^2\Sigma^+(100)$ state is repumped through the $\tilde{B}$ state, which has a measured branching ratio of $3\times10^{-3}$ to $\tilde{X}^2\Sigma^+(01^10)$ \cite{kozyryev2019determination}, we expect approximately $\red{0.043 \times 0.003 \approx 1.3 \times 10^{-4}}$ decay via this mechanism. The remaining decay occurs directly from the $\tilde{A}$ state and can be explained by a combination of Renner-Teller and spin-orbit interactions which mix the $\tilde{A}^2\Pi_{1/2}(000)$ and $\tilde{A}^2\Pi_{1/2}(01^10)$ states. This loss is predicted to occur at the \red{$\sim 2$ $ \times 10^{-4}$} level.

Decay to the $\tilde{X}^2\Sigma^+(02^20)$ state has not previously been observed in CaOH but is due to a Renner-Teller mechanism that mixes vibronic states with $\Delta l = \pm 2$ and $\Delta \Lambda = \mp 2$, where $\Lambda$ is the projection of the electronic angular momentum onto the molecular symmetry axis. Noticing that a $\Pi$ state can take both $\Lambda = +1$ and $\Lambda = -1$, this means that a direct mixing between $\tilde{A}^2\Pi_{1/2}(000)$ and $\tilde{A}^2\Pi_{1/2}(02^20)$ is allowed since these states have quantum numbers $(\Lambda = \pm 1, l = 0) \leftrightarrow (\Lambda = \mp 1, l = \pm 2)$ \cite{Hirota}. From this mechanism we predict a branching ratio of \red{$\sim 1\times10^{-3}$}, which agrees to within a factor of 3 with the measured value. Note that given its vibrational angular momentum, this state has $\Delta$ vibronic character and the rotationally closed transition is $J'=1/2 \rightarrow N''=2$ $(J''=\frac{3}{2})$. Combining all repumped states we measure that unaddressed loss channels contribute to decay at the \red{$4.5(7) \times 10^{-4}$} level and limit the us to an average of \red{$2200^{+400}_{-300}$} scattered photons per molecule.

Given these measured branching ratios, we can model decay out of the cycling transition as a \red{Markov} process where the probability of remaining in the cycling transition after each photon scatter is \red{$P_\text{rem} = 0.99955(7)$}. Then, from the ratio of molecule numbers detected with and without application of lasers and magnetic fields,  $\eta = N_\text{with}/N_\text{without}$, we can determine the number of photons $N$ scattered by the molecules using the simple relation \red{$P_\text{rem}^N = \eta$}. This analysis assumes that the molecules are not pushed out of the detection region by the interaction light, which is true when the main cycling laser is red detuned. For 12.5 cm of interaction length with a maximum intensity of \red{7 mW/cm$^2$} in the experiment, the population remaining was \red{$\eta = 0.66$}, corresponding to \red{920} photons scattered. 

\begin{table}
\centering
\begin{tabular}{| c | c |}
\hline
Decay & Branching Ratio \\
\hline
$(000)$ & $\red{0.9539(21)}$ \\
$(100)$ & $\red{0.0429(20)}$ \\
$(200)$ & $\red{0.9(2) \times 10^{-3}}$ \\
$(02^00)$ & $\red{1.3(2) \times 10^{-3}}$ \\
$(02^20)$ & $\red{3.1(5) \times 10^{-4}}$ \\
$(01^10)$ & $\red{2.4(4) \times 10^{-4}}$ \\
Other & $\red{4.5(7) \times 10^{-4}}$ \\
\hline
\end{tabular}
\caption{Branching ratios for decay from the laser cooling scheme depicted in Fig 1 from the main text. States are labeled following the convention from the main text as $\tilde{X}^2\Sigma^+(v_1 v_2^l v_3)$. All decays are measured on the rotationally closed $J'=1/2 \rightarrow N''=1$ $(J''=\frac{1}{2},\frac{3}{2})$ transitions used for laser cooling.}
\label{tab:FCFs}
\end{table}

\begin{figure}[ht]
\begin{centering}
\includegraphics[width = .5 \textwidth]{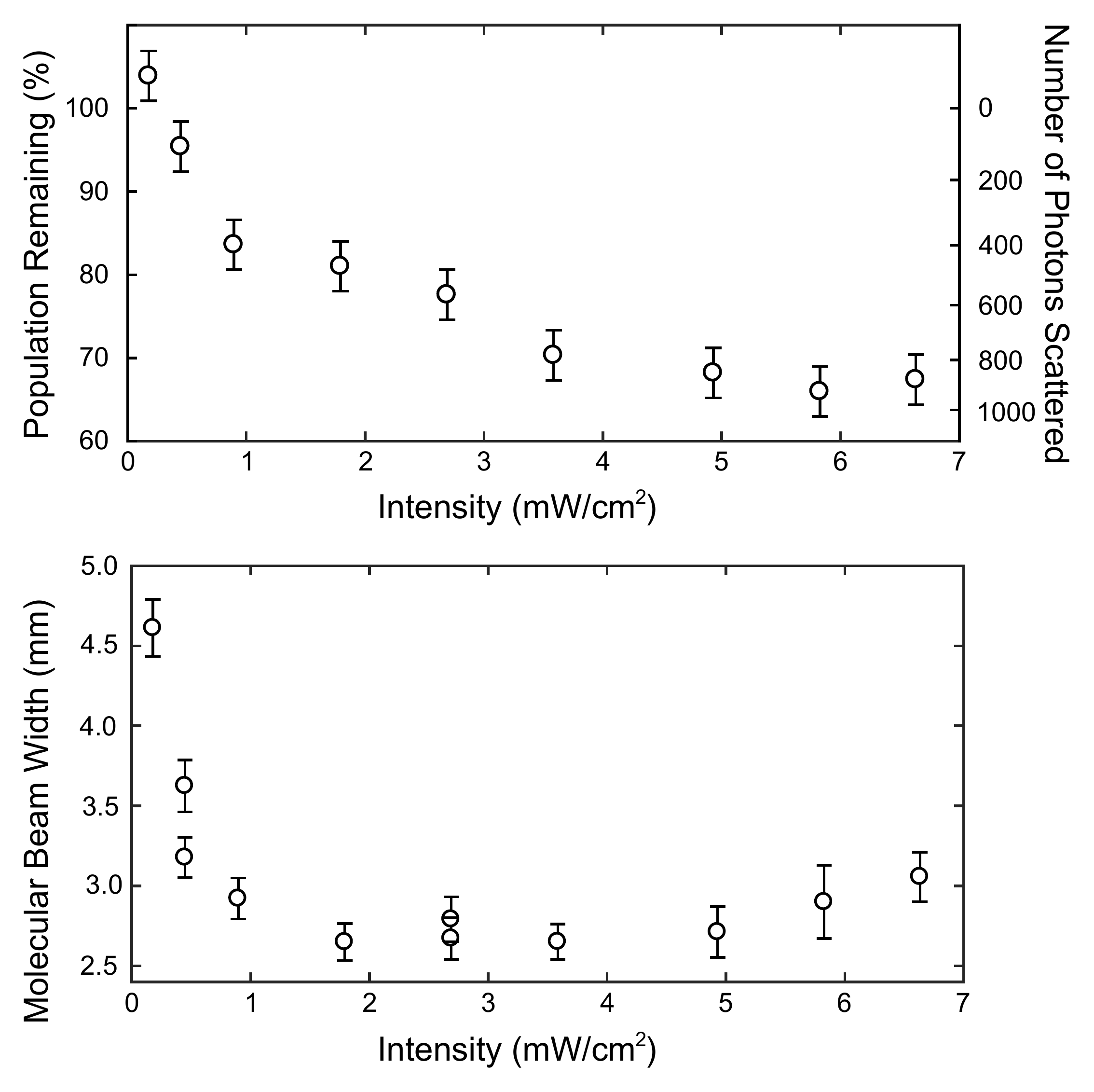} 
\caption{\label{fig:Widthvpower} Molecular beam width vs. power for the 1D MOT configuration. The main cooling laser was detuned -7 MHz and all repumping lasers were on resonance.}
\par\end{centering}
\vspace{-5mm}
\end{figure}

Fig. \ref{fig:Widthvpower} shows the width of the molecular beam as a function of applied main cooling laser intensity. The effect of the magneto-optical forces saturates at approximately \red{1.6 mW/cm$^2$}. We attribute the saturation to a combination of sub-Doppler heating and overfocusing of the molecular beam. At higher laser intensities the observed molecular beam width does not further narrow while the photon scattering rate nonetheless increases, resulting in a loss of population to unaddressed vibrational states. These two competing effects lead to a maximum enhancement of on-axis beam brightness at a laser intensity of \red{1.6 mW/cm$^2$}.

\begin{center}
\textbf{CaOH Zeeman Structure and MO Polarizations}
\end{center}

\begin{figure}[h]
\begin{centering}
\includegraphics[width = .5 \textwidth]{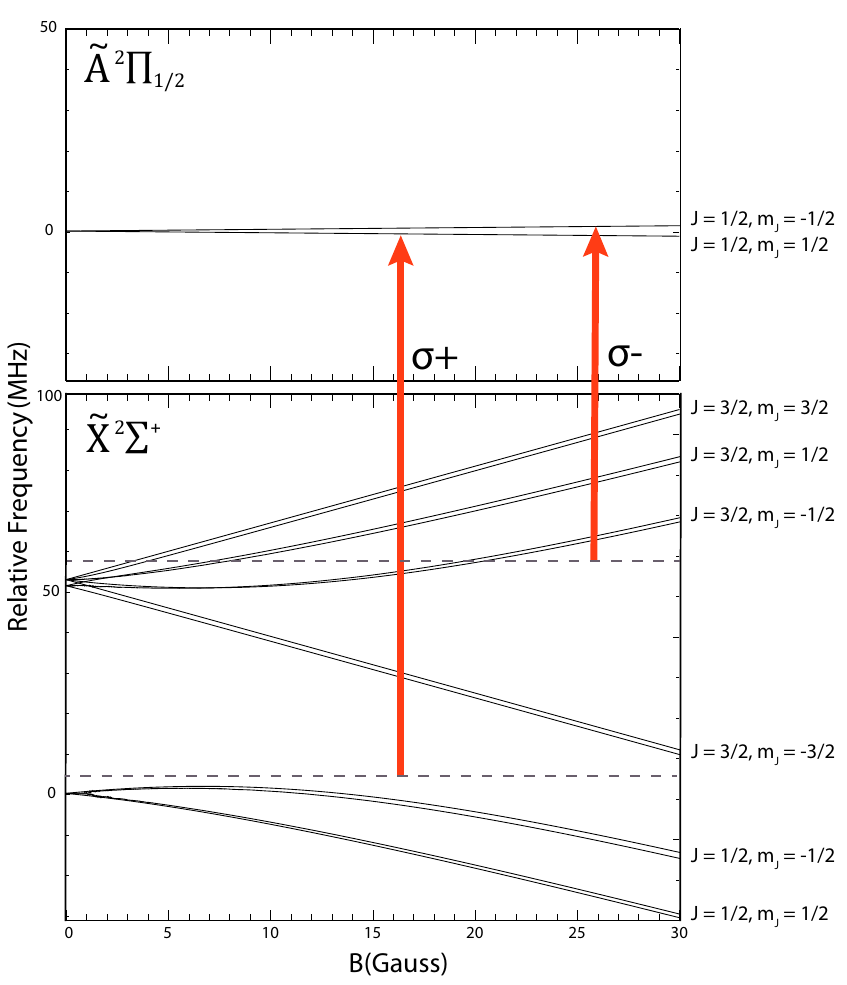} 
\caption{\label{fig:Zeeman}Plot of the calculated Zeeman shift of the ground and excitated states of CaOH. The much smaller g-factor in the excitated state is accentuated by the identical scale of the top and bottom panels.}
\par\end{centering}
\vspace{-5mm}
\end{figure}

Because the Land\'{e} $g$-factor of the $\tilde{A}^2\Pi_{1/2}(J'=\frac{1}{2})$ state in CaOH, $g_J^{(A)} \approx -0.021$, is more than an order of magnitude smaller than the $g$-factors of the $\tilde{X}^2\Sigma^+(J''=\frac{1}{2},\frac{3}{2})$ states, we rely on the Zeeman shift of the lower states to apply significant magneto-optical forces. The Zeeman structure of both electronic states is shown in Fig \ref{fig:Zeeman}. We label states in the $|J,m_J\rangle$ basis as hyperfine structure is significantly mixed at very small fields, making $m_J$ a good quantum number. At low fields $\lesssim 5$ G the Zeeman shifts in the ground state are approximately linear with $g$-factors $g_{J=1/2}^{(X)} = -2/3$ and $g_{J=3/2}^{(X)} = +2/3$, but deviate from linearity at fields $\gtrsim 10$ G due to the relatively small spin-rotation splitting of 52 MHz. Because the lower ($J''=1/2$) manifold has a negative $g$-factor while the upper ($J''=3/2$) manifold has a positive $g$-factor, the laser polarizations used to apply MO restoring forces have opposite handedness for the two manifolds, as depicted in Fig \ref{fig:Zeeman}.
The  $J \rightarrow J-1$ nature of this transition leads to rapid optical pumping of population into magnetic darks states. As descrbed in the main text, we remix these dark states by switching the sign of the magnetic field in conjunction with the handedness of the polarization. This switching occurs at $\sim$ 10$^6$ s$^{-1}$ which is on the order of our scattering rate.

\begin{center}
\textbf{Experimental Apparatus Details}
\end{center}

The RF magnetic field is generated by a pair of in-vacuum coils. These coils consist of 8 turns of copper with dimensions 21 mm $\times$ 164 mm directly bonded to an aluminum nitride substrate for thermal stability. The coils are mounted with a spacing of 23 mm. These coils are connected to independent resonant circuits outside the vacuum chamber which are driven by two radio frequency amplifiers. The inductance of the coils themselves is $\sim$~15~$\mu$H. The resonant tank circuits are designed to maximize the transfer of RF power from the amplifiers to the coils. They consist of two vacuum variable, high power capacitors (150-1500~pF, 4~kV and 70-1600~pF, 3~kV) in parallel with high power ceramic capacitors. One of the capacitors ($C_T$) is used to tune the resonance frequency while the other ($C_M$) is tuned to achieve impedance matching with the amplifier. For operation at 530~kHz, the nominal values of the capacitances are $C_M \approx$ 1000~pF and $C_T \approx$ 6000~pF. We employ fast electro-mechanical relays in order to switch the field on and off in $\sim$~10~ms while withstanding high RF powers.     

The magnitude and phase of the current through the MOT coils is monitored in situ by AC current probes. Under typical conditions the coils are driven at 530 kHz with  $I_{rms}$ = 9.6 A that corresponds to a magnetic field gradient of $B_{rms}$ = 17 Gauss/cm. These coils are operated at a duty cycle of 10\% to avoid heating and subsequent outgassing as seen in similar experiments.

\begin{center}
\textbf{Rate Equation Model and Beam Simulation}
\end{center}

We model the Doppler cooling and magneto-optical forces in our system using a rate-equation model similar to that described in \cite{Norrgard2016}. As shown in \cite{Tarbutt2013}, for a generic molecule with $N_g$ ground states coupled to $N_e$ excited states, and under the assumption that all excited states undergo spontaneous decay at the same rate $\Gamma$, the scattering rate can be expressed as
\begin{equation}
R_\text{sc} = \Gamma \frac{N_e}{(N_g + N_e) + 2\sum_{j=1}^{N_g}(1+4\Delta_j^2/\Gamma^2)I_{\text{sat},j}/I_j}
\end{equation}
where the sum is over all ground states $j$, $I_j$ is the laser intensity addressing the $j$th state, $\Delta_j$ is its detuning, and $I_{\text{sat},j} = \pi h c \Gamma/(3\lambda_j^3)$ is the corresponding two-level saturation intensity. Here $h$ is the Planck constant, $c$ is the speed of light, and $\lambda_j$ is the wavelength of the $j$th transition. In the case of CaOH, the $\tilde{X}^2\Sigma^+$(000) ground state has $N_g = 12$ hyperfine components and the $\tilde{A}^2\Pi_{1/2}$(000) excited state has $N_e = 4$ hyperfine states, $\Gamma = 2\pi \times 6.4$ MHz, and $\lambda = 626$ nm.

We further simplify this model by noting that all saturation intensities are approximately equal, $I_{\text{sat},j} \equiv I_\text{sat} = \pi h c \Gamma/(3\lambda^3)$, and that all transitions are equally detuned, $\Delta_j \equiv \Delta$. This latter condition arises because the two SR components in the ground state are individually addressed by splitting the laser into two frequencies with an acousto-optic modulator (AOM), and the laser is then globally detuned from resonance. The hyperfine splitting is small enough to be ignored at this level of approximation. Under these additional assumptions, the scattering rate due to a single laser simplifies to an effective two-level expression,
\begin{equation}
R_\text{sc}(\Delta) = \frac{\Gamma_\text{eff}}{2} \frac{s_\text{eff}}{1+s_\text{eff}' + 4\Delta^2/\Gamma^2}
\label{eqn:TwoLevelRsc}
\end{equation}
where
\begin{equation}
\Gamma_\text{eff} = \frac{2 N_e}{N_g + N_e} \Gamma
\end{equation}
is the effective scattering rate and
\begin{align}
\label{supp:approx}
s_\text{eff} &= \frac{N_g + N_e}{2 I_\text{sat}} \left( \frac{N_{g}^{(1/2)}}{I_{1/2}} + \frac{N_{g}^{(3/2)}}{I_{3/2}} \right)^{-1} \nonumber \\
&\approx \frac{N_g + N_e}{2 N_g} \frac{(I/2)}{I_\text{sat}}
\end{align}
is the effective saturation parameter. Here we assume that the total laser intensity addressing all $N_g^{(1/2)} = 4$ states in the $J''=1/2$ ground-state manifold is $I_{1/2}$, and light with intensity $I_{3/2}$ addresses the $J''=3/2$ manifold with $N_g^{(3/2)} = 8$ states 52 MHz away. Because of this large frequency separation, there is no cross talk between the two spin-rotation components. The approximation in Eq. \ref{supp:approx} assumes the total laser intensity $I$ is balanced between the two SR components and bears out the intuition that each state should be approximately resonant with half the total laser intensity in this case. The quantity $s_\text{eff}' \equiv I'/I_\text{sat}$ in the denominator of eqn. \ref{eqn:TwoLevelRsc} reflects the total intensity $I' > I$ of \emph{all} cooling lasers and is included to account for saturation effects. In an $n$-dimensional cooling scheme $s_\text{eff}'$ would equal $2ns_\text{eff}$.

We implement a Monte Carlo simulation of the MO forces in our CaOH beam by propagating $\sim 10^4$ molecules though a force field defined by the standard 1D expression $F_\text{MO} = \hbar k [R_\text{sc}(\Delta_1) - R_\text{sc}(\Delta_2)]$, where $k = 2\pi/\lambda$ and the detunings $\Delta_{1,2}$ account for laser detuning $\delta_0$ as well as Doppler and Zeeman shifts. The Zeeman shifts, in particular, are given by $\mu_\text{eff} A' r /\hbar$, where r is the radial distance from the magnetic field origin, $A' = (2\sqrt{2}/\pi)A_\text{rms}\cos\varphi$ is the time-averaged magnetic field gradient, $A_\text{rms}$ is the sinusoidal rms gradient applied in the experiment, and $\varphi$ is the phase offset between the oscillating gradient and laser polarization. $\mu_\text{eff}$ is the effective transition magnetic moment averaged over all 12 ground states (the excited state has negligible Zeeman shift); we estimate it as $\mu_\text{eff} \approx \mu_B/2$ by solving multi-level rate equations \cite{Tarbutt2015} for the steady-state population of each ground state. To account for the radial symmetry of the field gradient we project the polarization of each cooling laser onto the local quantization axis of every molecule and treat $\sigma^-$, $\pi$, and $\sigma^+$ polarization components as driving effective transitions with negative, zero, and positive magnetic moment, respectively.

The spatial and velocity distribution of the molecules incident on the cooling region is determined by initializing $\sim10^6$ molecules with a Gaussian spatial distribution of radial extent $\sigma_r \sim 8$mm and transverse temperature $T \sim 1$K at the output of the buffer gas cell. They ballistically propagate with Gaussian forward velocity distribution $v_f \sim 100\pm35$ m/s through the 3x3 mm beam aperture 35.5 mm downstream, after which only $\sim 0.1\%$ of the molecules remain, with a transverse temperature $T \sim 8.4$mK. Cooling forces are applied over a 12.5 cm length by five spatially separated, 25 mm diameter laser beams with uniform intensity. After cooling, the molecules ballistically propagate to the detection region, where their transverse spatial and velocity distributions are fit to Gaussians to extract beam width and temperature.

In order to fit the simulation to our data, we scale the overall MO force by making the replacement $\Gamma_\text{eff} \rightarrow \Gamma_\text{eff}' \equiv \zeta \Gamma_\text{eff}$ in the above equations, and we also take $\mu_\text{eff}$ as a fit parameter. All other constants are held fixed at the values used in the experiment. For optimal experimental parameters of $I$ = 1.6 mW/cm$^2$, $\delta_0 = -7$ MHz, and $A_\text{rms} = 17$ G/cm, we find that our simulation fits the data when $\zeta \approx 0.21$ and $\mu_\text{eff} \approx 0.53 \mu_B$. This value of $\mu_\text{eff}$ is in line with that predicted from multi-level rate equations as described above, and suggests that our spatial MO forces are approximately optimal. The factor of $\sim 5$ reduction in overall cooling rate parametrized by $\zeta$, meanwhile, is in line with similar imperfections previously quoted for a MOT of diatomic molecules \cite{truppe2017molecules}.

To extract the damping rate and MO oscillation frequency we linearize the applied force to find $F_\text{MO} \approx -\alpha v - \kappa r$, where $v$ and $r$ are the velocity and position of the molecules with respect to the magnetic field origin and the force constants are
\begin{align}
\alpha &= - \frac{8\hbar k^2 s_\text{eff} \Gamma_\text{eff}' \delta_0}{\Gamma^2(1+s_\text{eff}'+4\delta_0^2/\Gamma^2)^2} \\
\kappa &= \frac{\mu_\text{eff} A' \alpha}{\hbar k} = \frac{2\sqrt{2}}{\pi}\frac{\mu_\text{eff} A_\text{rms} \alpha}{\hbar k}\cos\varphi
\end{align}
where $\delta_0$ is the cooling laser detuning. Substituting the results of the fit described above yields a damping constant $\beta = \alpha/m \approx 400$ s$^{-1}$ and a MO frequency $\omega = \sqrt{\kappa/m} \approx 2\pi \times 90$ Hz, as quoted in the main text. Here $m$ is the mass of CaOH.

To estimate the capture velocity of a 3D MOT we use the full MO force profile described above with $\zeta$ and $\mu_\text{eff}$ fit to our data and use the same experimental parameters as in the fit. From this force profile, we numerically integrate the 1D equations of motion for a single molecule entering the 3D MOT along the molecular beam axis with variable initial velocity. The capture velocity is determined by finding the maximum initial velocity at which the molecular trajectory turns around (i.e. reaches negative forward velocity) before exiting the MOT volume defined by the cooling beams. We assume a 3D MOT beam diameter of 20 mm and uniform intensity. This results in an estimated on-axis capture velocity of $\lesssim 7$ m/s. Monte Carlo simulations performed in other work \cite{chae2017one} suggest that this value will be reduced by a factor of $\sim 2$ when averaged over all molecules, which in general enter the 3D MOT away from the central axis and experience different magnetic field and laser intensity profiles.

\end{document}